\documentclass[10pt]{article} 

\input amssym.def
\input amssym
\input epsf.sty

\newtheorem{theorem}{{\bf Theorem} \rm}[section]
\newtheorem{lemma}{{\bf Lemma} \rm}[section]

\newtheorem{claim}{{\bf Claim} \rm}[section]

\newtheorem{definition}{{\bf Definition} \rm}[section]

\newtheorem{fact}{{\bf Fact} \rm}[section]

\newenvironment{proof}{{\bf Proof:  }}{\qquad\rule{2mm}{2mm}}

\newcommand\union\cup
\newcommand\meet\wedge
\newcommand\implies\Rightarrow

\newcommand{\prob}[1]{\mbox{{\bf Pr}$\left[#1\right]$}}
\newcommand{\size}[1]{\left|#1\right|}
\newcommand{\ceil}[1]{\left\lceil#1\right\rceil}

\newcommand{\ket}[1]{\left|#1\right\rangle}
\newcommand{\norm}[1]{\left\|\,#1\,\right\|}

\def\bbbc{{\mathchoice {\setbox0=\hbox{$\displaystyle\rm C$}\hbox{\hbox
to0pt{\kern0.4\wd0\vrule height0.9\ht0\hss}\box0}}
{\setbox0=\hbox{$\textstyle\rm C$}\hbox{\hbox
to0pt{\kern0.4\wd0\vrule height0.9\ht0\hss}\box0}}
{\setbox0=\hbox{$\scriptstyle\rm C$}\hbox{\hbox
to0pt{\kern0.4\wd0\vrule height0.9\ht0\hss}\box0}}
{\setbox0=\hbox{$\scriptscriptstyle\rm C$}\hbox{\hbox
to0pt{\kern0.4\wd0\vrule height0.9\ht0\hss}\box0}}}}

\def\cents{\hbox{\rm\rlap/c}}

\newcommand{\ignore}[1]{}
\newcommand{\eq}{\;=\;}

\newcommand{\enc}[3]{{#1} \stackrel{{#3}}{\mapsto} {#2}}
\def\prob{{\rm Prob}}
\def\thmm{{Theorem}}

\newcommand{\inceps}[2]{
\epsfxsize= #2 cm
\centerline{\epsffile{#1.eps}}
}

\newcommand{\eqdef}{\stackrel{\rm def}{=}} 

\newcommand{\remove}[1]{}

\begin{document}

\title{Dense Quantum Coding and a Lower Bound for 1-way Quantum Automata}
\author{
{\em Andris Ambainis} 
\thanks{Computer Science Division, UC~Berkeley.
Email: {\tt ambainis@cs.berkeley.edu}.  
Supported by Berkeley Fellowship for Graduate Studies.}
\and
{\em Ashwin Nayak} 
\thanks{Computer Science Division, UC~Berkeley.
Email: {\tt ashwin@cs.berkeley.edu}.  
Supported by JSEP grant FDF 49620-97-1-0220-03-98.}
\and
{\em Amnon Ta-Shma} 
\thanks{International Computer Science Institute.
Email: {\tt amnon@icsi.berkeley.edu}. } 
\and
{\em Umesh Vazirani} 
\thanks{Computer Science Division, UC~Berkeley.
Email: {\tt vazirani@cs.berkeley.edu}. 
Supported by JSEP grant FDF 49620-97-1-0220-03-98.}
}
\date{}

\maketitle

\begin{abstract}
\noindent\normalsize
We consider the possibility of encoding~$m$ classical bits into 
much fewer~$n$ quantum bits so that an arbitrary bit from the original~$m$ 
bits can be recovered with a good probability,
and we show that non-trivial quantum encodings exist that have no
classical counterparts.
On the other hand, we show that quantum encodings cannot be much more
succint as compared to classical
encodings, and we provide a lower bound on such quantum encodings.
Finally, using this lower bound, we prove an exponential lower bound on
the size of 1-way quantum finite automata for a family of languages
accepted by linear sized deterministic finite automata.
\end{abstract}

\section{Introduction}

The tremendous information processing capabilities of
quantum mechanical systems may be attributed to the
fact that the state of an~$n$ quantum bit (qubit) system 
is given by a unit vector in a~$2^n$ dimensional complex
vector space. Since~$2^n - 1$ complex numbers are necessary to 
completely specify the state of~$n$ quantum bits, it may appear that
it is possible to encode a lot of information into it.
Nonetheless,
a fundamental result in quantum information
theory---Holevo's theorem~\cite{ho}---states
that no more than~$n$ classical bits of information 
can be transmitted by transferring~$n$ quantum bits from one
party to another. In view of this 
result, it is tempting to conclude that the exponentially
many degrees of freedom latent in the description of a 
quantum system must necessarily stay hidden or 
inaccessible. 

However, the situation is more subtle since in
quantum mechanics, the recipient of the~$n$ qubits has
a choice of measurements he can make to extract information
about their state. In general, these measurements do not 
commute. Thus making a particular measurement will, in 
general, disturb the system, thereby destroying some or all
the information that would have been revealed by another 
possible measurement. This opens up the possibility of 
quantum {\em random access\/} encodings.
Say we wish to encode~$m$ classical 
bits~$b_1 \cdots  b_m$ into~$n$ quantum bits~$(m \gg n)$. Then 
a quantum {\em random access\/} encoding with parameters~$m, n, p$
(or simply an~$\enc{m}{n}{p}$ encoding) consists of an encoding map 
from~$\{0,1\}^m$ to $\bbbc^{2^n}$, together with a sequence 
of~$m$ possible measurements for the recipient. If the 
recipient chooses the~$i$th measurement
and applies it to the encoding of~$b=b_1 \ldots b_m$, 
the result of the 
measurement is~$b_i$ with probability at least~$p$. 

\begin{definition}
A~$\enc{m}{n}{p}$ {\em random access\/} encoding is a function~$f:
\{0,1\}^m \times R \mapsto \bbbc^{2^n}$
such that for every~$1 \le i \le m$,
there is a 
measurement~${\cal O}_i$ that returns~$0$ or~$1$ and has the property
that
$$ \forall b \in \{0,1\}^m : ~ 
\prob(~ {\cal O}_i \ket{f(b,r)} = b_i ~) \;\ge\; p. $$ 
We call~$f$ the {\em encoding function}, and~${\cal O}_i$ the
{\em decoding functions}. 
\end{definition}

Notice that given the~$n$ qubits corresponding to a random access
encoding of some~$m$ bits, the recipient cannot simply 
make all~$m$ measurements and retrieve the encoded bits
(thus violating Holevo's bound), since any measurement
disturbs the state vector.
{\it A priori}, there is no reason to rule out the existence of 
a~$\enc{c^n}{n}{p}$ encoding for constants $c > 1$, $p > 1/2$.
In fact, even though~$\bbbc^k$ can accommodate
only~$k$ mutually orthogonal unit vectors, it can accommodate~$c^k$
almost mutually orthogonal unit vectors (i.e. vectors such that
the dot product of any two has absolute value less than~$1/10$, say).
This might lead one to believe that such encodings exist.
If such quantum random access encodings were possible, 
it would be possible to, for instance, encode the contents of 
an entire telephone directory in a few quantum bits such that the
recipient of these qubits could, via a suitably chosen measurement,
look up any {\em single\/} telephone number of his choice.

The main question that we consider in this paper is: for what values 
of~$m$,~$n$ and~$p$ do~$\enc{m}{n}{p}$ encodings exist? 
For classical encodings, where we encode~$m$ classical
bits into~$n$ classical bits, we know the answer. Let, for~$p \in
[0,1]$,~$H(p) = -p\log p - (1-p)\log(1-p)$ denote the {\em binary entropy
function}. We show:
\begin{theorem}
\label{thm-classical}
For any~$p > 1/2$, there exist~$\enc{m}{n}{p}$ classical encodings
with~$n = (1-H(p))m + O(\log m)$, and any~$\enc{m}{n}{p}$ classical encoding
has $n \ge (1-H(p))m$.
\end{theorem}

We then show that quantum encodings are more powerful than
classical encodings. On the one hand,
we show that no classical encoding can encode two
bits into one bit with decoding success probability greater than~$0.5$,
and on the other hand, we exhibit a~$\enc{2}{1}{0.85}$ quantum encoding.
In fact, as Ike Chuang~\cite{i} has shown, it is possible to encode~$3$
bits into~$1$ qubit with success probability~$\approx 0.79$
by taking advantage of the fact that the amplitudes in quantum states
can be complex numbers.
The~$2$-into-$1$ quantum encoding and the~$3$-into~$1$ encoding
easily generalize to a~$\enc{2n}{n}{0.85}$ and a~$\enc{3n}{n}{0.79}$ 
encoding, respectively. However, 
the question as to whether quantum encodings can {\em asymptotically\/}
beat the classical lower bound of Theorem~\ref{thm-classical} 
is left open.
Our main result about quantum encodings is that they cannot be much smaller
than the encoded strings.

\begin{theorem}
\label{thm:lb:quantum}
If a~$\enc{m}{n}{p}$ 
quantum encoding exists
with~$p> {1 \over 2}$ a constant, 
then~$n \ge \Omega({m \over {\log m}})$.
\end{theorem}
Thus, even though quantum random access encodings can beat
classical encodings, they cannot be much more succint.

We finish the paper with 
a novel application of our bound to showing a lower bound on the 
size of 1-way quantum finite automata (QFAs).
(See Section~\ref{sec-defn-qfa} for a precise definition of 1-way QFAs.)
In~\cite{kw} it was shown that
not every language recognized by a (classical)
deterministic finite automaton~(DFA)
can be recognized by a 1-way QFA.
On the other hand, 
there are languages that can be recognized
by 1-way QFAs with size exponentially smaller than that of corresponding
classical automata~\cite{af}. 
It remained open whether, for any language
that can be recognized by a 1-way finite automaton both classically and
quantum-mechanically, we can efficiently simulate
the classical automaton by a 1-way QFA. 
Our result answers 
this question in the negative, and demonstrates that while in some cases one
is able to exploit quantum phenomena to construct 
highly space-efficient 1-way
QFAs, in others, as it will become apparent, the requirement of 
the unitarity~(or, in other words, reversibility) of
evolution seriously limits their efficiency.

\begin{theorem}
\label{thm-qfa}
Let~$\{L_n\}_{n \ge 1}$ be a family of languages defined
by~$L_n=\{wa \;|\; w \in \{a,b\}^*, |w| \le n\}$.
Then, 
\begin{enumerate}
\item
\label{enu:det}
$L_n$ is recognized by a 1-way deterministic
automaton of size~$O(n)$,
\item
\label{enu:qu-upper}
$L_n$ is recognized by some 1-way quantum finite
automaton, and,
\item
\label{enu:qu-lower}
Any 1-way quantum automaton recognizing~ $L_n$ with
some constant probability greater than~${1\over 2}$  
has~$2^{\Omega(n/\log n)}$
states.
\end{enumerate}
\end{theorem}

\section{The classical bounds}

We first prove a lower bound on the number of bits required for a {\em
classical\/} random access encoding, and then show that there are
classical encodings that nearly achieve this bound. Together, these
yield Theorem~\ref{thm-classical} of the previous section.

The proof of the lower bound involves the concepts of the {\em
Shannon entropy\/}~$S(X)$ of a random variable~$X$, the Shannon
entropy~$S(X|Y)$ of a random variable~$X$ {\em conditioned on\/} another
random variable~$Y$, and the {\em mutual
information\/}~$I(X:Y)$ of a pair of random variables~$X,Y$. For definitions
and basic facts involving these concepts,
we refer the reader to a standard text (such as~\cite{ct}) on
information theory.

\begin{theorem}
Let $1/2 <p \leq 1$. For any classical~$\enc{m}{n}{p}$ encoding,~$n
\ge (1-H(p))m$.
\end{theorem}

\begin{proof}
Suppose there is such a (possibly probabilistic) encoding~$f$.
Let~$X = X_1\cdots X_m$ be chosen unformly at random from~$\{0,1\}^m$,
and let~$Y = f(X) \in \{0,1\}^n$ be the corresponding encoding. Let~$Z$
be the random variable with values in~$\{0,1\}^m$ obtained by generating
the bits~$Z_1\cdots Z_m$ from~$Y$ using the~$m$ decoding functions.

The mutual informaion of~$X$ and~$Y$ is clearly bounded by the number of
bits in~$Y$, i.e.~$n$:
$$ I(X:Y) \;\le\; S(Y) \;\le\; n. $$
We show below that it is, in fact, lower bounded by~$(1-H(p))m$, thus
getting our lower bound.

Now,
$$ I(X:Y) \eq S(X) - S(X|Y) \eq m - S(X|Y). $$
But, using standard properties of the entropy function, we have
$$ S(X|Y) \;\le\; S(X|Z) \;\le\; \sum_{i = 1}^m S(X_i|Z) \;\le\; \sum_{i =
1}^m S(X_i|Z_i). $$
It is not difficult to
see that~$S(X_i|Z_i) \le H(p)$. It follows that~$S(X|Y)
\le H(p)m$, and that~$I(X:Y) \ge (1-H(p))m$, as we intended to show.
\end{proof}

We now give an almost matching upper bound:

\begin{theorem}
There is a classical~$\enc{m}{n}{p}$ encoding with~$n=(1-H(p))m+O(\log m)$
for any~$p > {1\over 2}$.
\end{theorem}

\begin{proof}
The encoding is trivial for~$p > 1-{1\over m}$.
We describe the encoding for~$p \le 1-{1\over m}$ below.

We use a code~$S\subseteq \{0, 1\}^m$ such that,
for every~$x\in\{0, 1\}^m$, there is a~$y\in\{0, 1\}^m$
within Hamming distance~$(1-p-{1\over m})m$.
It is known (see, e.g.,~\cite{co}) that there is such a code~$S$
of size
$$ |S| \eq 2^{(1-H(p+{1\over m})m+2\log m} \leq 2^{(1-H(p))m+4\log m}. $$
Let~$S(x)$ denote the codeword closest to~$x$.
One possibility is to encode a string~$x$ by~$S(x)$. This would give us an
encoding of the right size. Further,
for every~$x$, at least $(p+{1\over m})m$ out of
the~$m$ bits would be correct.
This means that the probability (over all bits~$i$) 
that~$x_i=S(x)_i$ is at least~$p+1/m$. However, for our encoding
we need this probability to be at least~$p$ for {\em every\/} bit, 
not just on average over all bits. This can be achieved with the following
modification.

Let~$r$ be an~$m$-bit string,
and~$\pi$ be a permutation of~$\{1, \ldots, m\}$.
For a string~$x\in\{0, 1\}^m$, let~$\pi(x)$
denote the string~$x_{\pi(1)} x_{\pi(2)} \cdots x_{\pi(m)}$.

We consider encodings~$S_{\pi, r}$ 
defined by~$S_{\pi, r}(x) = \pi^{-1}(S(\pi(x+r))+r$.
We show that if~$\pi$ and~$r$ are chosen uniformly at random, then for any~$x$
and any index~$i$, the probability that the~$i$th bit in the encoding is
different from~$x_i$ is at most~$1-p-1/m$. First, note that
if~$i$ is also chosen uniformly at random, then this probability is 
clearly bounded by~$1-p-1/m$. So all we need to do is to show that this
probability is independent of~$i$.

If~$\pi$ and~$r$ are uniformly random, then~$\pi(x+r)$ is uniformly random
as well. Furthermore, for a fixed~$y=\pi(x+r)$, 
there is exactly one~$r$ corresponding to any permutation~$\pi$ 
that gives~$y=\pi(x+r)$.
Hence, if we condition on~$y = \pi(x+r)$,
all~$\pi$ (and, hence, all~$\pi^{-1}(i)$) are
equally likely. This means that the
probability that~$x_i \neq S_{\pi, r}(x)_i$ (or, equivalently,
that~$\pi(x+r)_{\pi^{-1}(i)} \neq (S(\pi(x+r))_{\pi^{-1}(i)}$) for
random~$\pi$ and~$r$ is just the probability of~$y_j\neq S(y)_j$
for random~$y$ and~$j$. This is clearly independent of~$i$ (and~$x$).

Finally, we show that there is a small set of permuation-string pairs such
that the desired
property continues to hold if we choose~$\pi,r$ uniformly at random
from {\em this\/} set, rather than the entire space of
permutations and strings. We
employ the probabilistic method to prove the existence of such a small set
of permutation-string pairs.

Let~$\ell =m^3$, and
let the strings~$r_1, \ldots, r_\ell \in \{0, 1\}^m$ and 
permutations~$\pi_1, \ldots, \pi_\ell$ be chosen
independently and uniformly at random.
Fix~$x\in\{0, 1\}^m$ and~$i\in [1..m]$.
Let~$X_j$ be~1 if~$x_i\neq S_{\pi_{j}, r_{j}}(x)_i$ and~0 otherwise.
Then~$\sum_{j=1}^\ell {X_j}$ is a sum of~$\ell$ independent Bernoulli
random
variables, the mean of which is at most~$(1-p-1/m)\ell$.
Note that~${1\over \ell} \sum_{j=1}^\ell {X_j}$ is the probability of
encoding the~$i$th bit of~$x$ erroneously when the permutation-string pair is 
chosen uniformly at random from the set~$\{(\pi_1,r_1),\ldots
(\pi_\ell,r_\ell) \}$.
By the Chernoff bound, the probability that the sum~$\sum_{j=1}^\ell
{X_j}$ is at
least~$(1-p-1/m)\ell+m^2$ (i.e., that the error probability~${1\over \ell}
\sum_{j=1}^\ell {X_j}$ mentioned above is at
least~$1-p$) is bounded by~${\rm e}^{-2m^4/\ell}={\rm e}^{-2m}$.
Now, the union bound implies that
the probability that the~$i$th bit of~$x$ is encoded erroneously with
probability more than~$1-p$ for
{\em any\/}~$x$ or~$i$ is at most~$m2^m{\rm e}^{-2m} < 1$.
Thus, there is a combination of strings~$r_1, \ldots, r_\ell$
and permutations~$\pi_1, \ldots, \pi_\ell$ with the property we seek.
We fix such a set of~$\ell$ strings and permutations.

We can now define our random access code as follows. To encode~$x$,
we select~$j\in\{1, \ldots, \ell\}$ uniformly at random and
compute~$y=S_{\pi_j, r_j}(x)$. This is the encoding of~$x$.
To decode the~$i$th bit, we just take~$y_i$.
For this scheme, we need $\log(\ell|S|)=\log \ell+\log |S|=(1-H(p))m+7\log
m$ bits. This completes the proof of the theorem.
\end{proof}

\section{A gap between quantum and classical encodings}

In this section, we show,
by exhibiting an encoding which has no classical
counterpart, that quantum encodings give us some advantage over
classical encodings.

\begin{figure}[t]
\label{fig:2to1}
\begin{center}
\inceps{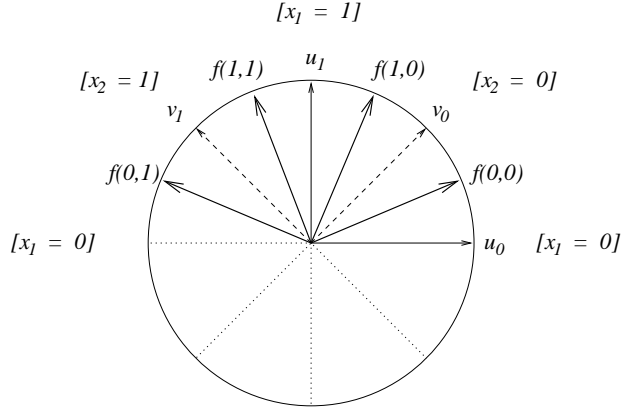}{8}
\caption{\it A~$2$-into-$1$ quantum encoding with probability of
success~$\approx 0.85$.}
\end{center}
\end{figure}

\begin{lemma}
There is a~$\enc{2}{1}{0.85}$ quantum encoding.
\end{lemma}

\begin{proof}
Let~$u_0=\ket{0}$, $u_1=\ket{1}$,
and~$v_0={1 \over \sqrt{2}} (\ket{1}+\ket{0})$, 
$v_1={1 \over \sqrt{2}} (\ket{1}-\ket{0})$.
Define~$f(x_1,x_2)$, the encoding of the string~$x_1 x_2$
to be~$u_{x_1}+v_{x_2}$ normalized
(See Figure~1).
The decoding functions are defined as follows:
for the first bit~$x_1$,
we measure the message qubit according
to the~$u$ basis and associate~$u_0$ with~$x_1=0$ and~$u_1$ with~$x_1=1$.
Similarly, for the second bit, we measure
according to the~$v$ basis,
and associate~$v_0$ with~$x_2=0$ and~$v_1$ with~$x_2=1$. 

It is easy to verify that for all four codewords, and for any~$i=1,2$,
the angle between the codeword and the right subspace is~$\pi/8$.
Hence the success probability is $\cos^2(\pi/8) \approx 0.853$.
\end{proof}

\begin{lemma}
\label{lem:no}
No~$\enc{2}{1}{p}$ classical encoding exists for any $p> {1 \over 2}$.
\end{lemma}

\begin{proof} 
Suppose there is a classical~$\enc{2}{1}{p}$ encoding
for some~$p>{1 \over 2}$.
Let~$f: \{0,1\}^2 \times R \mapsto \{0,1\}$ be the corresponding
probabilistic encoding
function and~$V_i : \{0,1\} \times R' \mapsto \{0,1\}$
the probabilistic decoding functions. 
If we let~$y_i$ be the random variable~$V_i (f(x,r),r')$, then
for any~$x \in \{0,1\}^2$, and any $i \in \{1,2\}$,
$\prob_{r,r'} ( y_i=x_i) \ge p$.

We first give a geometric characterization of the decoding functions. 
Each~$V_i$ clearly depends
only on the encoding, which is either~$0$ or~$1$. 
Define the point~$P^j$ (for~$j=0,1$) in the unit square~$[0,1]^2$ 
as~$P^j=(a^j_0,a^j_1)$, where~$a^j_i=\prob_{r'} (~V_i(j,r')=1~)$.
The point~$P^0$ characterizes the decoding functions when 
the encoding is~$0$, and~$P^1$ characterizes the decoding functions
when the encoding is~$1$. 
For example,~$P^1=(1,1)$ means that given the encoding~$1$,
the decoding functions return~$y_1=1$ and~$y_2=1$
with certainty, and~$P^0=(0,1/4)$ means that given the encoding~$0$,
the decoding functions return~$y_1=0$ and, with probability~$1/4$,
that~$y_2=1$.

\begin{figure}[t]
\label{fig:convex}
\begin{center}
\epsfxsize=2.4in
\hspace{0in}
\epsfbox{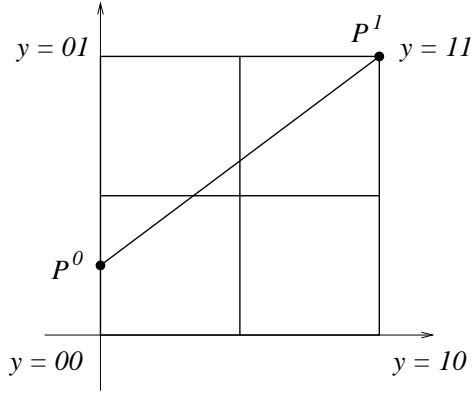}
\caption{\it A geometric characterization of the probabilistic decoding
functions of Lemma~\ref{lem:no}.}
\end{center}
\end{figure}

Any string~$x = x_1 x_2 \in \{0,1\}^2$ is encoded as a~$0$ with some
probability~$p_x$ and as a~$1$ with some probability~$1-p_x$.
If we let~$P^x=(a^x_0,a^x_1)$, where~$a^x_i$ is the probability that~$y_i=1$,
then~$P^x = p_x P^0 + (1-p_x) P^1$.
Thus,~$P^x$ lies on the line connecting the two points~$P^0$ and~$P^1$.
On the other hand, for the encoding to be a valid~$2$-into-$1$ encoding,
the point~$P^x$ should lie {\em strictly\/} inside the quarter of the unit
square~$[0,1]^2$ closest to~$(x_1,x_2)$.

Now, the line connecting~$P^0$ and~$P^1$
intersects the interiors of only three of the four 
quarters of the unit square~$[0,1]^2$. For instance, if~$P^0$
and~$P^1$ are as above, then the line
connecting them does not pass through the lower right quarter 
(see Figure~2).
Thus, for the string~$x_1 x_2$ which is favored by that quarter
(e.g. the string~$x=10$ in the example above),
either~$V_1$ or~$V_2$ errs with probability at least a half---which is a
contradiction.
\end{proof}

\section{The quantum lower bound}

We now prove \thmm~\ref{thm:lb:quantum}.
We first show that the success probability of the decoding process can be
amplified at the cost of a small increase in the length of the random
access code.

\begin{lemma}
\label{lem:amplify}
If for a constant~$p > {1 \over 2}$ there is an~$\enc{m}{n}{p}$ encoding,
then there is also an~$\enc{m}{O(n \log{1\over \epsilon})}{1-\epsilon}$
encoding for any~$\epsilon=\epsilon(m) > 0$.
\end{lemma}

\begin{proof}
Suppose there is an encoding~$f: \{0,1\}^m \times R \mapsto
\bbbc^{2^n}$ with 
decoding algorithms~${\cal O}_i$ ($i=1,\ldots,m$) with success
probability~$p > 1/2$.
We define a new encoding~$f^{(t)} : \{0,1\}^m \times R^t \mapsto
(\bbbc^{2^n})^{t}$ as~$f^{(t)}(x,r_1,\ldots,r_t)= f(x,r_1) \otimes
\cdots \otimes f(x,r_t)$.
I.e., it is the tensor product of~$t$ independent identical copies
of the original code.
The new decoding functions~${\cal O}_i'$ consists of
applying~${\cal O}_i$ to each of the~$t$ independent copies of
the code, and answering according to the majority.
The Chernoff bound shows that the error probability decays exponentially
fast in the number of trials, and is therefore at most~$\epsilon$
when~$t$ is chosen to be~$O(\log{1\over \epsilon})$.
\end{proof}

By choosing~$\epsilon = 1/q(m)$ for some polynomial~$q$,
we achieve an encoding with error~$\epsilon$ at
the cost of using an~$O(\log m)$ factor more qubits for the 
encoding. Now the result of any measurement cannot 
perturb the state vector too much (i.e.\ by more than~$\sqrt{\epsilon}$).
It might seem that this is sufficient to give us the lower bound, since 
we need to make only~$m$ measurements to recover all~$m$
encoded bits, and the error per measurement is only~$1/{\rm poly}(m)$.
However, the situation is more subtle, since 
the error on subsequent measurements must take into 
account both the encoding error, as well as the error 
introduced by previous measurements. In fact, a 
straightforward analysis suggests that the error doubles
with each measurement, thus making such a proof infeasible.
Instead, we prove that the errors grow linearly
(rather than exponentially), 
by first invoking the principle of safe storage (see~\cite{bv})
to defer all measurements to the end of a sequence of 
unitary operations, and then bounding the errors in the 
computation via a hybrid technique 
from~\cite{bbbv} (which is made more explicit in~\cite{v}).

\begin{lemma}
If a~$\enc{m}{n}{1-\epsilon}$ 
quantum encoding 
with~$\epsilon \le {1 \over {64 m^2}}$ exists,
then~$n \ge \Omega(m)$.
\end{lemma}

\begin{proof}
We first deal with {\em deterministic\/} quantum encoding, in which
the encoding function~$f: \{0,1\}^m \mapsto \bbbc^{2^n}$
maps inputs to {\em pure\/} states.
Any such encoding has, for every~$i \in [1..m]$, a decoding function
which takes a codeword~$\ket{\phi}$ and an ancilla~$\ket{0^l}$,
applies a unitary transformation~$V_i$, and makes a measurement.
Thus, it resolves~$\bbbc^{2^n}$ into two
subspaces~$W_i^0$ and~$(W_i^0)^\bot$ corresponding
to the answers~$0$ and~$1$ (for the $i$th bit), respectively.
Given~$\ket{\phi,0^l}$, we can thus decompose it 
as~$\ket{\phi_i^0} + \ket{\phi_i^1}$, where~$\ket{\phi_i^0} \in W_i^0$
and~$\ket{\phi_i^1} \in (W_i^0)^\bot$.

We now apply the principle of safe storage.
Instead of applying~$V_i$ and measuring,
we use unitary transformations~$U_i$ ($i=1,\ldots,m$) 
that work over the codeword~$\ket{\phi}$, 
the ancilla~$\ket{0^l}$ 
and~$m$ output bits~$\ket{0^m}$, such that~$U_i \ket{\phi_i^0,a}
= \ket{\phi_i^0,a}$ and~$U_i \ket{\phi_i^1,a} =
\ket{\phi_i^1,a \oplus e_i}$,
where~$e_i$ is the vector~$\ket{0,\ldots,0,1,0,\ldots0}$
having a~$1$ entry only in the~$i$th place.

The transformations~$U_i$ introduce some garbage at each step,
and their composition~$U_1 \cdots U_m$ is quite messy. To analyse their
behaviour, we first fix an input~$x$,
and imagine ideal unitary transformations~$U_i'=U_i'(x)$
that have the property that for the codeword~$\ket{\phi_x}$
of~$x$,~$U_i' \ket{\phi_x,a} = \ket{\phi_x,a \oplus (x_i \cdot e_i)}$.
Since for any~$x \in \{0,1\}^m$ and any~$i \in [1..m]$,
the transformation~$U_i$ correctly yield the~$i$th bit of~$x$ with high
probability, the reader can verify that
\begin{eqnarray}
\label{eq:onestep}
\norm{U_i \ket{\phi_x,0^l,a} - U_i' \ket{\phi_x,0^l,a} }^2
& \le &
2\epsilon.
\end{eqnarray}
We now claim that the result of applying the transformations~$U_i$ does
not differ
much from that of applying the ideal transformations~$U'_i$.

\begin{claim}
$\norm{ U_1 \cdots U_m \ket{\phi_x,0^l,0^m}  -
        U_1' \cdots U_m' \ket{\phi_x,0^l,0^m} } \le 2m \sqrt{\epsilon}$.
\end{claim}
  
\begin{proof}
We use a hybrid argument:
\begin{eqnarray*}
\norm{ U_1 \cdots U_m \ket{\phi_x,0^l,0^m}  - 
       U_1' \cdots U_m' \ket{\phi_x,0^l,0^m} } & \le &
\norm{ U_1 \cdots U_{m-1} U_m \ket{\phi_x,0^l,0^m}  -  
       U_1 \cdots U_{m-1} U_m'  \ket{\phi_x,0^l,0^m} } +  \\
& &
\norm{ U_1 \cdots U_{m-1} U_m' \ket{\phi_x,0^l,0^m}  -  
       U_1 \cdots U_{m-1}' U_m'  \ket{\phi_x,0^l,0^m} } +  \\
& & \cdots + \\ 
& & \norm{ U_1  \cdots U_{m-1}' U_m' \ket{\phi_x,0^l,0^m}  -  
           U_1' \cdots U_{m-1}' U_m'  \ket{\phi_x,0^l,0^m} }   \\
\end{eqnarray*}

But, since the transformations~$U_i$ are unitary, we have:
\begin{eqnarray*}
\lefteqn{ \norm{U_1 \cdots U_t U_{t+1}'\cdots U_m' \ket{\phi_x,0^l,0^m}  - 
       U_1 \cdots U_t' U_{t+1}' \cdots U_m' \ket{\phi_x,0^l,0^m} } } \\
 & = & \norm{U_t U_{t+1}' \cdots U_m' \ket{\phi_x,0^l,0^m}  - 
       U_t' U_{t+1}' \cdots U_m' \ket{\phi_x,0^l,0^m} } \\
 & = & \norm{U_t \ket{\phi_{t+1}'} - U_t' \ket{\phi_{t+1}'} },
\end{eqnarray*}
where~$\ket{\phi_{t+1}'} = U_{t+1}' \cdots U_m' \ket{\phi_x,0^l,0^m}$.
By the definition of the transformations~$U'_i$,~$\ket{\phi_{t+1}'} = 
 \ket{\phi_x,0^l,a}$ with~$a=\ket{0,\ldots,0,x_{t+1},\ldots,x_m}$.
Hence, by equation~(\ref{eq:onestep}),
$\norm{U_t \ket{\phi_{t+1}'} - U_t' \ket{\phi_{t+1}'} } \le 2 \sqrt{\epsilon}$,
and the claimed result follows.
\end{proof}

Now we can extract all the bits of~$x$ by computing~$\ket{\psi}= 
U_1 \ldots U_m \ket{\phi_x,0^l,0^m}$
and measuring the~$m$ answer bits~$a_1,\ldots,a_m$. The following claim
says that we succeed with high probability.

\begin{claim}
$\prob( a \ne x) \le 4m \sqrt{\epsilon}$.
\end{claim}

\begin{proof}
Let~$\ket{\psi'}=U_1' \ldots U_m' \ket{\phi_x,0^l,0^m} = \ket{\phi_x,0^l,x}$.
From the claim above,
we know that~$\norm{\ket{\psi}-\ket{\psi'}} \le 2m \sqrt{\epsilon}$.
When we measure the answer bits of~$\ket{\psi'}$, we get~$x$ with
probability~1.
Moreover, from the following fact, the probability of observing~$x$ on
measuring~$\ket{\psi}$ cannot differ from this by very much.
\begin{fact}
Suppose~$\norm{\ket{\psi_1}-\ket{\psi_2}} \le \delta$.
Let~${\cal O}$ be a measurement with possible results~$\Lambda$,
and~${\cal D}_i$ the classical distributions over~$\Lambda$
that result from applying~${\cal O}$ to~$\ket{\psi_i}$.
Then $\norm{{\cal D}_1-{\cal D}_2}_1 \eqdef 
\Sigma_{a \in \Lambda} |{\cal D}_1(a)-{\cal D}_2(a)| \le 2\delta$.
\end{fact}
Hence, the probability that~$a \neq x$ is at most $4m\sqrt{\epsilon}$.
\end{proof}

Therefore, we get~$x$ with probability at least~$1-4m \sqrt{\epsilon}
\ge 1- {4m \over 8m} = {1 \over 2}$.
It then follows from Holevo's Theorem~\cite{ho} that~$n \ge \Omega(m)$.

Now we deal with 
{\em probabilistic\/} quantum 
encoding, where we can encode a string~$x \in \{0,1\}^m$
as a probabilistic mixture of pure states.
It is well known that we can always {\em purify\/} the system,
i.e., we can adjoin ancilla bits to the encoding, such that the result is 
a pure state.
Now, as before, we may apply the decoding 
transformations~$U_i$ and retrieve all the encoded bits: for every~$x$,
there are ideal transformations~$U_i'=U_i'(x)$ that behave almost as~$U_i$
(in the same sense as above) and the same argument again gives us the
lower bound on~$n$.
\end{proof}

Combining the two lemmas above, we get \thmm~\ref{thm:lb:quantum}.
We remark that we may extend this lower bound to general~$p > 1/2$, by
appropriately generalizing Lemma~\ref{lem:amplify} above.

\subsection{Serial encodings}

We note that \thmm~\ref{thm:lb:quantum} holds even in a slightly 
more general scenario, when the decoding functions are allowed to depend
on the string encoded.

\begin{definition}
$f: \{0,1\}^m \times R \mapsto {\cal C}^{2^n}$
{\em serially encodes\/}~$m$ classical bits into~$n$ qubits
with~$p$ success, 
if for any~$i \in [1..n]$ and~$b_{[i+1,n]} = b_{i+1}\cdots b_n
\in\{0,1\}^{n-i} $,
there is a 
measurement~${\cal O}_{i,b_{[i+1,n]}}$ 
that returns~$0$ or~$1$ and has the property that
$$ \forall b \in \{0,1\}^m: ~
\prob
( ~{\cal O}_{i,b_{[i+1,n]}} \ket{f(b,r)}  ~=~b_i ~) \ge p.
$$ 
\end{definition}

I.e., we allow the decoding functions to depend on the
suffix~$b_{i+1}\cdots b_m$ of the string~$b$ for
recovering the value of the~$i$th bit~$b_i$. The lower bound for quantum
random access codes of the previous section also holds for serial
encodings.

\begin{theorem}
\label{thm-serial}
Any quantum serial encoding of~$m$ bits into~$n$ qubits with
constant success probability~$p > {1 \over 2}$
has~$n \ge \Omega({m \over {\log m}})$.
\end{theorem}

\begin{proof}
On careful examination, we see that for
the proof of Theorem~\ref{thm:lb:quantum} to work in this case as well,
all we need to check is that for all~$i \in [1..n]$,
\begin{eqnarray*}
\norm{U_i \ket{\phi_x,0^l,a_i} - U_i' \ket{\phi_x,0^l,a_i} }^2
& \le &
2\epsilon,
\end{eqnarray*}
where~$a_i=\ket{0,\ldots,0,x_{i+1},\ldots,x_m}$.
Although the transformations~$U_i$ may now depend on the bits already
decoded, the above bound is easily verified, since~$a_i$ contains the required
suffix of the encoded word~$x$.
\end{proof}

\section{The lower bound for 1-way quantum finite automata}
\label{sec-lb-qfa}

In this section, we give the details of the proof of
Theorem~\ref{thm-qfa}.
The first two parts of \thmm \ref{thm-qfa} are easy.
Figure~3
shows a DFA with~$2n+3$ states for the language~$L_n$.
Also, Since each~$L_n$ is a finite language, there is
a 1-way {\em reversible\/} finite automaton (as defined in
Section~\ref{sec-defn-qfa}), and hence a 1-way QFA that accepts it.
What then remains to be shown is the lower bound on the size of 
a 1-way QFA accepting the language.

\begin{figure}[t]
\label{fig-dfa}
\begin{center}
\epsfxsize=6.4in
\hspace{0in}
\epsfbox{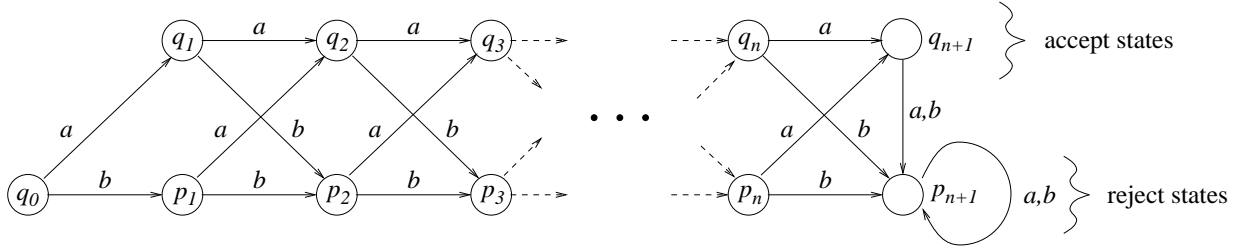}
\caption{\it A DFA that accepts the language~$L_n = \{ wa \;|\; w \in
\{a,b\}^*, \size{w} \le n \}$.}
\end{center}
\end{figure}

Intuitively, since a 1-way QFA
is allowed to read input symbols only once, a QFA for~$L_n$ 
necessarily ``records'' the last
symbol read in its state, and since it is required to be reversible, it
is forced to ``remember'' {\em all} the symbols read until it is clear
whether the input is in the language or not. Thus, we expect 
the state of the automaton after~$n$ input symbols to be an
{\em encoding\/} of the~$n$ symbols.
It is not difficult to see that in the case of a 1-way reversible automaton
that accepts the language~$L_n$, the encoding is such that all the~$n$ input
symbols can be recovered with certainty. Thus, such an automaton 
has at least~$2^n$ states. 
However, for reasons stated below, it is not clear in the
case of a {\em general\/} 1-way QFA that the state encodes the input
symbols in a ``faithful'' manner. 
\begin{itemize}
\item
Firstly, a 1-way QFA is allowed to make {\em partial
decisions\/} (i.e., it is allowed to accept or reject an input with some
probability before reading all its symbols). We show in
Section~\ref{sec-ext} that partial decisions can be ``deferred'' for~$r$
steps at a cost of only an~$O(r)$ factor increase in the size of the
automaton. We call the resulting automaton an~$r$-{\em restricted\/}
QFA. Since no input of length more than~$n+1$ belongs to~$L_n$,
this means that partial decisions are not very useful in building
``small'' automata for the language, and that we can limit our study to
that of~$n$-restricted QFAs.
\item
Secondly, and more seriously,
the encoding defined by the
automaton is such that each input symbol is accessible via a 
measurement only when all the symbols following it are known,
and by trying to learn the later symbols we might destroy
the encoding.
 
This problem is exactly the one \thmm~\ref{thm-serial}
solves. We can thus conclude that the number of qubits required to
represent a state of the automaton is~$\Omega(n/\log n)$, which gives
us the lower bound stated in Theorem~\ref{thm-qfa}.
\end{itemize}

Before presenting the formal proof for the lower bound, we define
1-way QFAs precisely in the next section. We then show how a restricted QFA
for the language~$L_n$ yields a
serial encoding of~$n$ classical bits into a state of the automaton.
Theorem~\ref{thm-serial} then immediately gives a size lower bound
of~$2^{\Omega(n/\log n)}$ for restricted QFAs. We then extend this lower
bound to general QFAs in Section~\ref{sec-ext}.

\subsection{Technical preliminaries}
\label{sec-defn-qfa}

A 1-way quantum finite automaton~(QFA) is a theoretical
model for a quantum computer with finite memory.
It has a finite set of basis states~$Q$, which consists of three parts:
accepting states, rejecting states and non-halting states.
The sets of accepting, rejecting and non-halting basis states are
denoted by~$Q_{\rm acc}, Q_{\rm rej}$ and $Q_{\rm non}$, respectively.
One of the states,~$q_0$, is distinguished as the starting state.

Inputs to a QFA are words over a finite alphabet~$\Sigma$.
We shall also use the symbols~`$\cents$' and~`$\$$'
that do not belong to~$\Sigma$ to 
denote the left and the right end marker, respectively.
The set~$\Gamma = \Sigma\cup\{\cents, \$, \}$ denotes the working alphabet
of the QFA.
For each symbol~$\sigma\in\Gamma$, a 1-way QFA has a corresponding unitary
transformation~$U_{\sigma}$ on the space~$\bbbc^Q$.
A 1-way QFA is thus defined by describing~$Q, Q_{\rm acc},
Q_{\rm rej}, Q_{\rm non}, q_0, \Sigma$, and~$U_\sigma$ for
all~$\sigma\in\Gamma$. We will often refer to 1-way QFAs as simply
QFAs, since we do not consider any other type of QFAs in this paper.

At any time, the state of a QFA is a superposition
of basis states in~$Q$. The computation starts in the 
superposition~$|q_0\rangle$. Then transformations corresponding to 
the left end marker~`$\cents$,' the letters of the input word~$x$ and
the right end marker~`$\$$' are applied in succession to the state of
the automaton, unless a transformation results in acceptance or rejection
of the input. A transformation corresponding to a symbol~$\sigma\in\Gamma$
consists of two steps:
\begin{enumerate}
\item
First, $U_\sigma$ is applied to~$\ket{\psi}$, the current state of the
automaton, to obtain the new state~$\ket{\psi'}$.
\item
Then, $\ket{\psi'}$ is measured with respect to 
the observable~$E_{\rm acc}\oplus E_{\rm rej}\oplus E_{\rm non}$,
where $E_{\rm acc}={\rm span}\{|q\rangle \;|\; q\in Q_{\rm acc}\}$,
$E_{\rm rej}={\rm span}\{|q\rangle \;|\; q\in Q_{\rm rej}\}$,
$E_{\rm non}={\rm span}\{|q\rangle \;|\; q\in Q_{\rm non}\}$.
The probability of observing~$E_i$ is
equal to the squared norm of the projection of~$\ket{\psi'}$ onto~$E_i$.
On measurement, the state of the automaton ``collapses'' to the projection
onto the space observed, i.e., becomes equal to the projection, suitably
normalized to a unit superposition.

If we observe~$E_{\rm acc}$ (or~$E_{\rm rej}$), the input is accepted
(or rejected). Otherwise, the computation continues, and the 
next transformation, if any, is applied.
\end{enumerate}
We regard these two steps together as reading the symbol~$\sigma$.

A QFA~$M$ is said to {\em accept\/} (or {\em recognize}) a language~$L$ with
probability~$p > {1\over 2}$ if 
it accepts every word in~$L$ with probability at least~$p$, 
and rejects every word not in~$L$ with probability at least~$p$.

A {\em reversible finite automaton\/}~(RFA) is a QFA such that,
for any $\sigma\in\Gamma$ and $q\in Q$, $U_{\sigma}\ket{q}=\ket{q'}$
for some $q'\in Q$. In other words, the operator~$U_\sigma$ is a
permutation over the basis states; it maps each basis state to a basis
state, not to a superposition over several states.

The {\em size\/} of a finite automaton is defined
as the number of (basis) states in it. The ``space used by the
automaton'' refers to the number of (qu)bits required to represent an
arbitrary automaton state.

\subsection{The lower bound for restricted QFAs}
\label{sec-enc}

Define an {\em $r$-restricted\/} 1-way QFA for a language~$L$ 
as a 1-way QFA that recognizes the language with probability~$p > {1\over 2}$,
and which halts with non-zero probability before seeing the right end marker 
only {\em after\/} it has read~$r$ letters of the input.
We first show a lower bound on the size of $n$-restricted 1-way QFAs that
accept~$L_n$.

Let~$M$ be any $n$-restricted 1-way QFA accepting~$L_n$ with constant
probability~$p > {1 \over 2}$. The following claim formalizes the intuition 
that the state of~$M$ after~$n$ symbols of the input have been read is
an encoding of the input string.
\begin{claim}
There is a serial encoding of~$n$ bits into~$\bbbc^Q$, and hence
into~$\ceil{\log\size{Q}}$ {\em qubits}, where~$Q$ is the set of basis
states of the QFA~$M$.
\end{claim}
\begin{proof}
Let~$Q$ be the set of basis states of the QFA~$M$,
and let~$Q_{\rm acc}$ and~$Q_{\rm rej}$ be the
set of accepting and rejecting states respectively.
Also, let~$U_\sigma$ be the unitary operator of~$M$
corresponding to the symbol~$\sigma \in \{a,b,\cents, \$\}$.
Let~$E_{\rm acc}, E_{\rm rej}$ and~$E_{\rm non}$ be defined as in
Section~\ref{sec-defn-qfa}.

We define an encoding~$f : \{a,b\}^n \rightarrow \bbbc^Q$
of~$n$-bit strings into unit superpositions over the basis states of the
QFA~$M$ by letting~$\ket{f(x)}$ be the state of the automaton~$M$ after
the input string~$x\in \{a,b\}^n$ has been read. We assert that~$f$ is
a serial encoding.

To show that~$f$ is indeed such an encoding, we exhibit a suitable
measurement for the~$i$th bit of the input for every~$i \in [1..n]$.
Let, for~$y\in \{a,b\}^{n-i}$, $V_i(y) = U_\$
U^{-1}_y$, where~$U_y$ stands for the identity operator if~$y$ is the
empty word, and for~$U_{y_{n-i}} U_{y_{n-i-1}} \cdots U_{y_1}$ otherwise.
The~$i$th measurement then consists of first
applying the unitary transformation~$V_i(x_{i+1}\cdots x_n)$ to~$\ket{f(x)}$,
and then measuring the resulting superposition with respect
to~$E_{\rm acc} \oplus E_{\rm rej} \oplus E_{\rm non}$.
(Note that the measurement for the~$i$th bit assumes the knowledge of
all the successive bits~$x_{i+1},\ldots,x_n$ of the input.)
Since for words with length at most~$n$, containment in~$L_n$ is decided
by the last letter, and because such words are accepted or rejected by
the~$n$-restricted QFA~$M$ with probability at least~$p$
{\em only after the entire input has been read},
the probability of observing~$E_{\rm acc}$ if~$x_i = a$,
or~$E_{\rm rej}$ if~$x_i = b$, is at least~$p$. Thus,~$f$ defines a
serial encoding, as claimed.
\end{proof}

Theorem~\ref{thm-serial} now immediately implies that~$\ceil{\log
\size{Q}} = \Omega(n/\log n)$ and thus~$\size{Q} = 2^{\Omega(n/\log
n)}$, where~$Q$ is as in the claim above. 

\subsection{Extension to general QFAs}
\label{sec-ext}

It only remains to show that the lower bound on the size of restricted QFAs
obtained above implies a lower bound on the size of general QFAs
accepting~$L_n$. We do this by showing that we can convert {\em any\/}
1-way QFA to an~$r$-restricted 1-way QFA which is only~$O(r)$ times
as large as the
original QFA. It follows that the~$2^{\Omega(n/\log n)}$ lower bound on
number of states of~$n$-restricted 1-way QFAs recognizing~$L_n$
continues to hold for general 1-way QFAs for~$L_n$, exactly as stated in
Theorem~\ref{thm-qfa}.

The idea behind the construction of a restricted QFA, given a general
QFA, is to carry the halting parts of the superposition of the original
automaton as ``distinguished'' non-halting parts of the state of 
the new automaton till at least~$r$ more symbols of the input have
been read since the halting part was
generated or until the right end marker is encountered, and then mapping
them to accepting or rejecting subspaces appropriately.

\begin{lemma}
Let~$M$ be a 1-way QFA with~$S$ states
recognizing a language~$L$ with probability~$p$.
Then there is an~$r$-restricted 1-way QFA~$M'$ with~$O(rS)$ states
that recognizes~$L$ with probability~$p$.
\end{lemma}
\begin{proof}
Let~$M$ be a 1-way QFA with~$Q$ as the set of basis states,~$Q_{\rm acc}$
as the set of accepting states,~$Q_{\rm rej}$ as the set of rejecting states,
and~$q_0$ as the starting state. Let~$M'$ be the automaton with basis state
set
$$ Q \;\union\; (Q_{\rm acc} \times \{0,1,\ldots,r+1\} \times \{{\rm
acc},{\rm non}\}) \;\union\; (Q_{\rm rej} \times \{0,1,\ldots,r+1\}
\times \{{\rm rej},{\rm non}\}). $$
Let~$Q_{\rm acc}\union (Q_{\rm acc} \times
\{0,1,\ldots,r+1\} \times \{{\rm acc}\})$ be its set of accepting
states, let~$Q_{\rm rej}\union (Q_{\rm rej} \times \{0,1,\ldots,r+1\} \times
\{{\rm rej}\})$ be the set of rejecting states, and let~$q_0$ be the 
starting state. If, for a state~$q \in Q$,
there is a transition
\begin{eqnarray*} 
\ket{q} & \mapsto & \sum_{q'} \alpha_{q'} \ket{q'}
\end{eqnarray*}
in~$M$ on symbol~$\sigma$, then in~$M'$, we have the following
transitions. On the~`\$' symbol,
we have the same transition, and on~$\sigma \not=
\$$, we have
\begin{eqnarray*} 
\ket{q} & \mapsto & \sum_{q'\not\in Q_{\rm acc}\union Q_{\rm rej}}
			  \alpha_{q'} \ket{q'} +
                          \sum_{q'\in Q_{\rm acc}\union Q_{\rm rej}}
			  \alpha_{q'} \ket{q',0,{\rm non}}.
\end{eqnarray*}
The transitions from the states not originally in~$M$ are given by the
following rules. On the~`\$' symbol,
\begin{eqnarray*} 
\ket{q,i,{\rm non}} & \mapsto & \left\{ \begin{array}{ll}
				\ket{q,i,{\rm acc}} & {\rm if\ } q \in
				Q_{\rm acc}\ {\rm and\ } i \le r \\
				     &     \\
				\ket{q,i,{\rm rej}} & {\rm if\ } q \in
				Q_{\rm rej}\ {\rm and\ } i \le r
				    \end{array}
			        \right.
\end{eqnarray*}
and on a symbol~$\sigma \in \{a,b\}$,
\begin{eqnarray*} 
\ket{q,i,{\rm non}} & \mapsto & \left\{ \begin{array}{ll}
				\ket{q,i+1,{\rm non}} & {\rm if\ } i < r
				                        \\
				 & \\
				\ket{q,i+1,{\rm acc}} & {\rm if\ } q
				\in Q_{\rm acc}\ {\rm and\ } i = r \\
				 & \\
				\ket{q,i+1,{\rm rej}} & {\rm if\ } q
				\in Q_{\rm rej}\ {\rm and\ } i = r
                                    \end{array}
			    \right.
\end{eqnarray*}
The rest of the transitions may be defined arbitrarily, subject to the
condition of unitarity.

It is not difficult to verify that~$M'$ is an~$r$-restricted 1-way QFA
(of size~$O(rS)$) accepting the same language as~$M$, and with the 
same probability.
\end{proof}

\subsection{Some remarks}
\label{sec-final}

We observe that the size~$O(n)$ versus
size~$\Omega(2^n)$ separation between DFAs and
1-way QFAs is the worst possible
if we restrict ourselves to languages that can be accepted by 1-way QFAs
with probability of correctness that is high enough (at least 7/9).
Such languages include all {\em finite} regular languages, since these
can be accepted by 1-way RFAs. This follows from the result
of Ambainis and Freivalds~\cite{af} that any language accepted by a QFA
with high enough probability can be
accepted by a 1-way RFA which
is at most exponentially bigger than the minimal DFA accepting the
language. However, it is not clear that this is also the largest separation
in the case of languages that are accepted by 1-way QFAs with smaller
probability of correctness.

Another open problem involves the blow up in size while
simulating a 1-way probabilistic finite automata~(PFA) by a 1-way QFA.
The only known way for doing this is by simulating the 
PFA by a 1-way DFA and then
simulating the DFA by a QFA. Both simulating a PFA
by a DFA~\cite{am,fr,ra} and simulating a DFA
by a QFA (this paper) can involve exponential or nearly exponential
increase in size. This means that the straightforward simulation of
a probabilistic automaton by a QFA (described above) could result in a
doubly-exponential increase in the size. However, we do not know of any 
examples where both transforming a PFA into a DFA and transforming
a DFA into a QFA cause big increases of size.
Better simulations of probabilistic automata by QFAs
may well be possible.

In general, it is not known how to simulate a probabilistic 
coin-flip by a purely quantum-mechanical algorithm if
space is limited. For example, the only known simulation
of $S(n)$-space probabilistic Turing machines by $S(n)$-space 
quantum Turing machines can create quantum Turing machines
running in expected time of $2^{2^{S(n)}}$\cite{wa}.
Finding better simulations or proving that they do not exist is
another interesting direction to explore.

\section*{Acknowledgements}
We would like to thank Ike Chuang for showing us the~$3$-into-$1$ quantum
encoding.
We also would like to thank Dorit Aharonov, Ike Chuang, Michael Nielsen,
Steven Rudich and Avi Wigderson for many interesting discussions.

\end{document}